\documentstyle[11pt,epsfig]{article}
\date{}
\begin{document}
\title{{\bf Noether symmetric classical and quantum scalar field cosmology}}
\author{Babak Vakili$^{1}$\thanks{b-vakili@iauc.ac.ir}\,\, and\,\,\ Farhad Khazaie
 $^{2}$ \,\,\,\\\\$^1${\small {\it Department of Physics, Chalous Branch of Azad
University, P.O. Box 46615-397, Chalous, Iran}}\\$^2${\small {\it
Department of Physics, Tehran Central Branch of Azad University,
Tehran, Iran}} $^{}${}} \maketitle

\begin{abstract}
We study the evolution of a two dimensional minisuperspace
cosmological model in classical and quantum levels by the Noether
symmetry approach. The phase space variables turn out to
correspond to the scale factor of a Friedmann-Robertson-Walker
(FRW) model and a scalar field with which the action of the model
is augmented. It is shown that the minisuperspace of such a model
is a two dimensional manifold with vanishing Ricci scalar. We
present a coordinate transformation which cast the corresponding
minisuper metric to a Minkowskian or Euclidean one according to
the choices of an ordinary or phantom model for the scalar field.
Then, the Noether symmetry of such a cosmological model is
investigated by utilizing the behavior of the corresponding
Lagrangian under the infinitesimal generators of the desired
symmetry. We explicitly calculate the form of the scalar field
potential functions for which such symmetries exist. For these
potential functions, the exact classical and quantum solutions in
the cases where the scalar field is an ordinary or a phantom one,
are presented and compared.\vspace{5mm}\noindent
\vspace{0.8mm}\newline Keywords: Noether symmetry, Scalar field
cosmology, Quantum cosmology
\end{abstract}
\section{Introduction}
Classical, semiclassical and quantum scalar fields have played a
central role in conceptual discussion of unified theories of
interactions and also in all branches of the modern cosmological
theories. From a cosmological point of view, there is a renewed
interest in the scalar-tensor models in which a non-minimal
coupling appears between the geometry of space-time and a scalar
field \cite{1}. This is because a number of different scenarios in
cosmology such as spatially flat and accelerated expanding
universe at the present time \cite{2}, inflation \cite{3}, dark
matter and dark energy \cite{4}, and a rich variety of behaviors
can be accommodated phenomenologically by scalar fields.
Traditionally, cosmological models of inflation use a single
scalar field with a canonical kinetic term of the form $1/2
g^{\mu\nu}\partial_{\mu}\phi\partial_{\nu}\phi$ with some
particular self-interaction potential $V(\phi)$ like $1/2m\phi^2$
or $\lambda \phi^4$, etc. Such a scalar field is often known as
minimally coupled to the geometry. However, there are also scalar
fields in what is qualified to be called the scalar-tensor theory,
which are not simply added to the tensor gravitational field, but
comes into play through the non-minimal coupling term \cite{5}. On
the other hand, the current cosmological observations allow the
possibility of existence of a cosmic fluid with equation of state
parameter smaller than $-1$ \cite{6}. One of the simplest
explanations of such an equation of state is considering a scalar
field with negative kinetic energy which is called a phantom field
\cite{7}. The classical and quantum cosmological dynamics of the
phantom scalar fields has been studied in a number of works, see
for example \cite{8}.

In this paper we shall study the classical and quantum dynamics of
a flat FRW model with a scalar field as its source by the Noether
symmetry approach. We set up an effective Lagrangian in which the
scale factor $a$ and scalar field $\phi$ play the role of
independent dynamical variables. This Lagrangian is constructed in
such a way that its variation with respect to $a$ and $\phi$
yields the appropriate equations of motion. The form of the
potential function of the scalar field is then found by demanding
that the Lagrangian admits the desired Noether symmetry \cite{9}.
By the Noether symmetry of a given minisuperspace cosmological
model we mean that there exists a vector field $X$, as the
infinitesimal generator of the symmetry, on the tangent space of
the configuration space such that the Lie derivative of the
Lagrangian with respect to this vector field vanishes. In
\cite{10} the applications of the Noether symmetry in various
cosmological models are studied. Here, before applying the Noether
symmetry condition on the model under consideration, we shall see
that the corresponding minisuperspace of our model is a two
dimensional Riemannian manifold with zero Ricci scalar. Therefore,
it is possible to find a new set of coordinates in terms of which
the minisuper metric takes the form of a Minkowskian (when the
scalar field is an ordinary quintessence field) or Euclidean (when
the scalar field is a phantom field) space. The desired coordinate
transformation brings the Lagrangian into a canonical form where
its kinetic and part of its potential terms are like the ones of a
couple of harmonic oscillators. We shall see by demanding the
Noether symmetry as a feature of this Lagrangian, we can obtain
the explicit form of the potential function. Since the existence
of a symmetry results in a constant of motion, we can integrate
the field equations which would then lead to expansion law of the
universe.

The structure of the article is as follows. In section 2, we
briefly present the basic elements of the issue of Noether
symmetry. In section 3, we first introduce the scalar field FRW
cosmology and write its Lagrangian in terms of the minisuperspace
variables. This section is then divided into two subsections each
of which will deal with one kind (quintessence or phantom) of
scalar field. In these two subsections, after applying the Noether
symmetry condition on the model, we provide the analytical
solutions for the corresponding Noether symmetric classical and
quantum cosmologies. Finally, the conclusions are summarized in
section 4.
\section{The Noether symmetry}
In this section we briefly review the issue of the Noether
symmetry approach used in the present work which originally
appeared in \cite{Cap}. To do this, we consider a dynamical system
with finite degrees of freedom moving in a Riemannian space with
the metric components ${\cal G}_{AB}$. The evolution of such a
system can be produced by the following action
\begin{equation}\label{A}
{\cal S}=\int_{{\cal M}} d\tau \left[\frac{1}{2}{\cal
G}_{AB}\frac{dq^A}{d\tau}\frac{dq^B}{d\tau}-{\cal U}(\bf
{q})\right],\end{equation}where $q^A$ are the dynamical variables
representing the configuration space of the system (the indices
$A$, $B$, ... run over the dimension of this space), ${\cal
U}(\bf{q})$ is the potential function and $\tau$ is an affine
parameter along the evolution path of the system. In time
parameterized theories, for instance general relativity, in which
their action is invariant under time reparameterization, the
affine parameter $\tau$ can be linked to a time parameter $t$ by a
lapse function $N(t)$ through $Ndt=d\tau$. Therefore, for such a
dynamical systems the action (\ref{A}) can be written as
\begin{equation}\label{B}
{\cal S}=\int_{{\cal M}}dt{\cal L}(q^A,\dot{q}^A)=\int_{{\cal
M}}dt N\left[\frac{1}{2N^2}{\cal G}_{AB}\dot{q}^A\dot{q}^B-{\cal
U}(\bf{q})\right],\end{equation}where now an over dot indicates
derivation with respect to the time parameter $t$ and ${\cal
L}(\bf{q},\dot{\bf{q}})$ is the Lagrangian function of the system.
Variation of the above action with respect to $q^A$ yields the
equations of motion as
\begin{equation}\label{C}
\frac{1}{N}\frac{d}{dt}\left(\frac{\dot{q}^A}{N}\right)+\frac{1}{N^2}\Gamma^A_{BC}\dot{q}^B\dot{q}^C=F^A,\end{equation}
where $F^A=-{\cal G}^{AB}\partial_{B}{\cal U}$ is a force term and
$\Gamma^A_{BC}=\frac{1}{2}{\cal G}^{AD}\left(\partial_B{\cal
G}_{DC}+\partial_C{\cal G}_{BD}-\partial_D{\cal G}_{BC}\right)$ is
the Christoffel symbols associated with the metric ${\cal
G}_{AB}$. In this formalism $N$ is not a dynamical variable in the
sense that variation of the action (\ref{B}) with respect to it
yields
\begin{equation}\label{D}
\frac{1}{2N^2}{\cal G}_{AB}\dot{q}^A\dot{q}^B+{\cal
U}(\bf{q})=0,\end{equation}which is nothing but the Hamiltonian
constraint, a constraint that should hold during the evolution of
the system and is indeed a reflect of the fact that the underlying
theory is a time parameterized theory. Basically, the equations of
motion (\ref{C}) are the geodesic equations of a "point particle"
(=the system) moving in a Riemannian space under the act of the
potential field ${\cal U}(\bf{q})$. Quantization of such a
dynamical system may be achieved by the method of canonical
quantization. For this purpose we introduce the momenta conjugate
to the variables $q^A$ by
\begin{equation}\label{E}
P_A=\frac{\partial {\cal L}}{\partial \dot{q}^A}={\cal
G}_{AB}\frac{\dot{q}^B}{N},\end{equation}leading to the following
Hamiltonian function
\begin{equation}\label{F}
H=P_A\dot{q}^A-{\cal L}=N\left[\frac{1}{2}{\cal
G}^{AB}P_AP_B+{\cal U}(\bf{q})\right]=N{\cal
H}.\end{equation}Because of the constraint equation (\ref{D}), the
above Hamiltonian seems to be identically equal to zero.
Therefore, under canonical quantization this Hamiltonian yields
the Wheeler-DeWitt (WDW) equation ${\cal H}\Psi({\bf q})=0$, where
$\Psi ({\bf q})$ is the wave function of the quantized system and
${\cal H}$ should be written in a suitable operator form. If one
makes a natural choice of factor ordering, the WDW equation may be
written as \cite{Dw}
\begin{equation}\label{G}
{\cal H}\Psi({\bf q})=\left[-\frac{1}{2}\nabla^2+{\cal U}({\bf
q})\right]\Psi({\bf q})=0,\end{equation}where
$\nabla^2=\frac{1}{\sqrt{-{\cal G}}}\partial_A\left(\sqrt{-{\cal
G}}{\cal G}^{AB}\partial_B\right)$ is the Laplacian operator in
the space with metric ${\cal G}_{AB}$.

Now, let us deal with the idea of the Noether symmetry in a given
dynamical system like that one presented above. Following
\cite{Cap}, we define the Noether symmetry as a vector field $X$
on the tangent space $TQ=({\bf q}, {\bf \dot{q}})$ of the
configuration space through
\begin{equation}\label{H}
X=\alpha^A({\bf q})\frac{\partial}{\partial
q^A}+\frac{d\alpha^A({\bf q})}{dt}\frac{\partial}{\partial
\dot{q}^A},\end{equation}where $\alpha^A({\bf q})$ are unknown
functions on configuration space. The Noether symmetry then
implies that the Lie derivative of the Lagrangian with respect to
this vector field vanishes, that is, $L_X{\cal L}=0$, which leads
\begin{equation}\label{I}
L_X{\cal L}=\alpha^A({\bf q})\frac{\partial {\cal L}}{\partial
q^A}+\frac{d\alpha^A({\bf q})}{dt}\frac{\partial {\cal
L}}{\partial \dot{q}^A}=0.\end{equation}Noether symmetry approach
is a powerful tool in finding the solutions to a given Lagrangian,
including the one presented above. In this approach, one is
concerned with finding the cyclic variables related to conserved
quantities and consequently reducing the dynamics of the system to
a manageable one. The existence of Noether symmetry means that
phase flux is conserved along the vector field $X$ and thus a
constant of motion exists. Indeed, noting that $P_A=\frac{\partial
{\cal L}}{\partial \dot{q}^A}$ and also taking into account the
Euler-Lagrange equations $\frac{dP_A}{dt}=\frac{\partial {\cal
L}}{\partial q^A}$, from (\ref{I}) we get
$\frac{d}{dt}\left[\alpha^A({\bf q})P_A\right]=0$. Thus the
constants of motion are found as
\begin{equation}\label{J}
Q=\alpha^A({\bf q})P_A.\end{equation}In order to obtain the
functions $\alpha^A({\bf q})$ we use equation (\ref{I}). However,
in some cases which we prefer to use the Hamiltonian formalism of
the theory, the Hamiltonian counterpart of this equation is better
suited in finding these coefficients which, equivalently, can be
written as
\begin{equation}\label{K}
\alpha^A\left\{P_A,H\right\}+\frac{\partial \alpha^A}{\partial
q^B}\left\{q^B,H\right\}P_A=0.\end{equation}In general, the
expression above gives a quadratic polynomial in terms of momenta
with coefficients being partial derivatives of $\alpha^A$ with
respect to the configuration variables ${\bf q}$. Thus, the
expression is identically equal to zero if and only if these
coefficients are zero, leading to a system of partial differential
equations for $\alpha^A({\bf q})$. An important ingredient in any
model theory related to Noether symmetry is the choice of the
cyclic variables related to this symmetry. Since such a variable
gives a constant of motion, it should have a vanishing commutator
with the Hamiltonian and therefore from a quantum mechanical point
of view they should have simultaneous eigenfunctions. This means
that to describe the quantum structure for a Noether symmetric
model, one should take into account the commutative algebra
between the Hamiltonian and the conserved quantities (\ref{J}). If
$Q$ is itself a cyclic variable the quantum counterpart of the
theory can be described by the following equations
\begin{eqnarray}\label{L}
\left\{
\begin{array}{ll}
{\cal H}\Psi({\bf q})=0,\\\\
\hat{Q}\Psi({\bf q})={\cal Q}\Psi({\bf q}),\\
\end{array}
\right.
\end{eqnarray}where $\hat{Q}$ is the operator form of (\ref{J})
and ${\cal Q}$ is its eigenvalue. If $Q$ is not a cyclic variable
this procedure does not work. In this case we seek a change of
variables in the form of a point transformation ${\bf
q}=(q^1,q^2,...)\rightarrow {\bf u}=(u^1,u^2,...)$ on the vector
field (\ref{H}) such that in terms of the new variables ${\bf u}$,
the Lagrangian includes one (or more) cyclic variable. A general
discussion of this issue can be found in \cite{Cap}. Under such a
point transformation it is easy to show that the vector field
(\ref{H}) takes the form \cite{Vak}
\begin{equation}\label{M}
\tilde{X}=(Xu^A)\frac{\partial}{\partial
u^A}+\frac{d}{dt}(Xu^A)\frac{\partial}{\partial
\dot{u}^A}.\end{equation}it is easy to show that if $X$ is a
Noether symmetry, $\tilde{X}$ has also this property, that is,
$X{\cal L}=0 \Rightarrow \tilde{X}{\cal L}=0$. Thus, if we demand
$Xu^i=1$ for some $u^i\in {\bf u}$ and $Xu^j=0$ for any $j\neq i$,
we get
\begin{equation}\label{N}
\tilde{X}=\frac{\partial}{\partial u^i}\Rightarrow \tilde{X}{\cal
L}=\frac{\partial {\cal L}}{\partial u^i}=0.\end{equation}This
means that $u^i$ is a cyclic variable and the dynamics can be
reduced. On the other hand, the constant of motion $Q$ becomes
\begin{equation}\label{O}
Q=\alpha^A({\bf q})P_A=\alpha^A({\bf q})\frac{\partial {\cal
L}}{\partial \dot{q}^A}=\alpha^A({\bf q})\left(\frac{\partial
{\cal L}}{\partial u^B}\frac{\partial u^B}{\partial
\dot{q}^A}+\frac{\partial {\cal L}}{\partial
\dot{u}^B}\frac{\partial \dot{u}^B}{\partial \dot{q}^A}\right).
\end{equation}Since ${\bf q}\rightarrow {\bf u}$ is a point transformation, we
have $\frac{\partial u^B}{\partial \dot{q}^A}=0$ and
$\frac{\partial \dot{u}^B}{\partial \dot{q}^A}=\frac{\partial
u^B}{\partial q^A}$, Therefore,
\begin{equation}\label{P}
Q=\alpha^A({\bf q})\frac{\partial u^B}{\partial q^A}\frac{\partial
{\cal L}}{\partial \dot{u}^B}=Xu^B\frac{\partial {\cal
L}}{\partial \dot{u}^B}=\frac{\partial {\cal L}}{\partial
\dot{u}^i}=\pi_i,\end{equation}where $\pi_i$ is the momentum
conjugate to $u^i$. Thus, as expected the constant of motion which
corresponds to the Noether symmetry is nothing but the momentum
conjugated to the cyclic variable. Now, it is obvious that
$\left[\hat{\pi_i},H\right]=0$ and the quantum description of the
system under consideration can be given by the following equations
\begin{eqnarray}\label{Q}
\left\{
\begin{array}{ll}
{\cal H}\Psi({\bf q})=0,\\\\
\hat{\pi_i}\Psi({\bf q})=\sigma_i\Psi({\bf q}),\\
\end{array}
\right.
\end{eqnarray}where $\sigma_i$ is the eigenvalue of
$\hat{\pi_i}$.

In the next section we will apply the above formalism to a given
cosmological model as a dynamical system. In cosmological systems,
since the scale factors, matter fields and their conjugate momenta
play the role of dynamical variables, introduction of Noether
symmetry by adopting the approach discussed above is particularly
relevant.
\section{Scalar field cosmology}
In this section we consider a FRW cosmology with a scalar field
with which the action of the model is augmented. In a
quasi-spherical polar coordinate the geometry of such a space-time
is described by the metric
\begin{equation}\label{R}
ds^2=-N^2(t)dt^2+a^2(t)\left[\frac{dr^2}{1-kr^2}+r^2(d\theta^2+\sin^2\theta
d\phi^2)\right],\end{equation}where $N(t)$ is the lapse function,
$a(t)$ the scale factor and $k=0,\pm 1$ is the curvature index.
Since our goal is to study the models which exhibit Noether
symmetry, we do not include any matter contribution in the action.
Let us start from the action (we work in units where
$c=\hbar=16\pi G=1$)
\begin{equation}\label{S}
{\cal S}=\int d^4x\sqrt{-g}\left[R-\frac{1}{2}\epsilon g^{\mu
\nu}\partial_{\mu}\phi\partial_{\nu}\phi-V(\phi)\right],\end{equation}where
$g$ is the determinant of the metric tensor $g_{\mu \nu}$, $R$ is
the Ricci scalar corresponding to $g_{\mu \nu}$, $V(\phi)$ is the
potential function for the scalar field $\phi(t)$ and the
parameter $\epsilon=\pm1$ corresponds to the ordinary scalar field
(where $\epsilon=+1$) or phantom scalar field (where
$\epsilon=-1$). By substituting (\ref{R}) into (\ref{S}) and
integrating over spatial dimensions, we are led to a point-like
Lagrangian in the minisuperspace $\{a,\phi\}$ as
\begin{equation}\label{T}
{\cal L}=-3a\dot{a}^2+\frac{1}{2}\epsilon
a^3\dot{\phi}^2+3ka-a^3V(\phi),\end{equation}in which we have set
$N=1$ so that the time parameter $t$ is the usual cosmic time.
Now, it is easy to see that this minisuperspace has the following
minisuper metric
\begin{equation}\label{U}
{\cal G}_{AB}dq^Adq^B=-6ada^2+\epsilon
a^3d\phi^2.\end{equation}This two dimensional minisuperspace is
the space spanned with the dynamical variables of the model and
indeed is the space in which we are looking for a Noether
symmetry. In the following we will apply the Noether symmetry
approach to the minisuperspace which is represented by a curved
manifold with a minisuper metric given by (\ref{U}).

\subsection{Ordinary scalar field}In this case ($\epsilon =+1$) the minisuperspace
has a Lorentzian metric. However, a simple calculation shows that
its Ricci scalar is zero and therefore this space is a two
dimensional Lorentzian flat space. This means that there is a
suitable change of variables that can transform the metric into a
Minkowskian one. To do this, consider the following change of
variables $q^A=(a,\phi)\rightarrow Q^A=(x,y)$ \cite{Bas}
\begin{equation}\label{V}
x=\frac{1}{\omega}a^{3/2}\cosh (\omega
\phi),\hspace{0.5cm}y=\frac{1}{\omega}a^{3/2}\sinh (\omega
\phi),\end{equation}where $\omega=\sqrt{6}/4$. In terms of these
new variables, Lagrangian (\ref{T}) takes the form
\begin{equation}\label{W}
{\cal
L}=\frac{1}{2}\left(\dot{x}^2-\dot{y}^2\right)-\frac{1}{2}\omega^2\left(x^2-y^2\right)
V\left(\frac{y}{x}\right)+3k\omega^{2/3}\left(x^2-y^2\right)^{1/3},\end{equation}with
the corresponding Hamiltonian becoming
\begin{equation}\label{X}
{\cal
H}=\frac{1}{2}\left(P_x^2-P_y^2\right)+\frac{1}{2}\omega^2\left(x^2-y^2\right)V\left(\frac{y}{x}\right)
-3k\omega^{2/3}\left(x^2-y^2\right)^{1/3},\end{equation}where
$P_x$ and $P_y$ are the momenta conjugate to $x$ and $y$
respectively. Thus, it is easy to see that in the minisuperspace
constructed by $Q^A=(x, y)$, the metric is Minkowskian and
represented by
\begin{equation}\label{Y}
\bar{{\cal G}}_{AB}dQ^AdQ^B=-dy^2+dx^2.\end{equation}

Now, we have a set of variables $(x, y)$ endowing the
minisuperspace with a Minkowskian metric and we are going to find
the potential function $V(y/x)$ such that the dynamical system
described with the Lagrangian (\ref{W}) exhibits the issue of the
Noether symmetry. For this purpose, following the steps after
equation (\ref{H}) we define the generator of the desired Noether
symmetry as the vector field
\begin{equation}\label{Z}
X=\alpha \frac{\partial}{\partial x}+\beta
\frac{\partial}{\partial y}+\dot{\alpha} \frac{\partial}{\partial
\dot{x}}+\dot{\beta} \frac{\partial}{\partial
\dot{y}},\end{equation}on the tangent space of the corresponding
configuration space such that the Lie derivative of the Lagrangian
with respect to this vector field vanishes, that is, $L_X{\cal
L}=0$. In (\ref{Z}) $\alpha$ and $\beta$ are some functions of $x$
and $y$ which in order to obtain them we use the Noether symmetry
condition. For Lagrangian (\ref{W}) this condition results in
\begin{eqnarray}\label{AB}
&& \alpha \left[-\omega^2 x
V(z)+\frac{1}{2}\omega^2\frac{y}{x^2}\left(x^2-y^2\right)V'(z)+2k\omega^{2/3}x\left(x^2-y^2\right)^{-2/3}\right]+
\nonumber\\&&  \beta \left[\omega^2 y
V(z)-\frac{1}{2}\omega^2\frac{1}{x}\left(x^2-y^2\right)V'(z)-2k\omega^{2/3}y\left(x^2-y^2\right)^{-2/3}\right]+\nonumber\\&&
\left(\frac{\partial \alpha}{\partial x}\dot{x}+\frac{\partial
\alpha}{\partial y}\dot{y}\right)\dot{x}- \left(\frac{\partial
\beta}{\partial x}\dot{x}+\frac{\partial \beta}{\partial
y}\dot{y}\right)\dot{y}=0,
\end{eqnarray}where $z=y/x$ and $V'(z)=dV/dz$. Now, we are led to
the following system of equations

\begin{equation}\label{AC}
\omega^2V(z)(\beta y-\alpha x)+\frac{1}{2}\omega^2
V'(z)\left(x^2-y^2\right)\left(\alpha \frac{y}{x^2}-\beta
\frac{1}{x}\right)+2k\omega^{2/3}\left(x^2-y^2\right)^{-2/3}(\alpha
x-\beta y)=0,\end{equation}
\begin{equation}\label{AD}
\frac{\partial \alpha}{\partial x}=\frac{\partial \beta}{\partial
y}=0,\hspace{0.5cm}\frac{\partial \alpha}{\partial
y}-\frac{\partial \beta}{\partial x}=0.\end{equation}Equations
(\ref{AD}) admit the independent solutions as
\begin{equation}\label{AE}
(\alpha,\beta)=(1,0),\hspace{0.5cm}(\alpha,\beta)=(0,1),\hspace{0.5cm}(\alpha,\beta)=(y,x),\end{equation}which
upon substitution into relation (\ref{AC}) yields the potential
function. In what follows we restrict ourselves to the flat case
$k=0$ and consider three cases separately.

\subsubsection{The case: $(\alpha,\beta)=(1,0)$} 
\begin{figure}
\begin{tabular}{c}\epsfig{figure=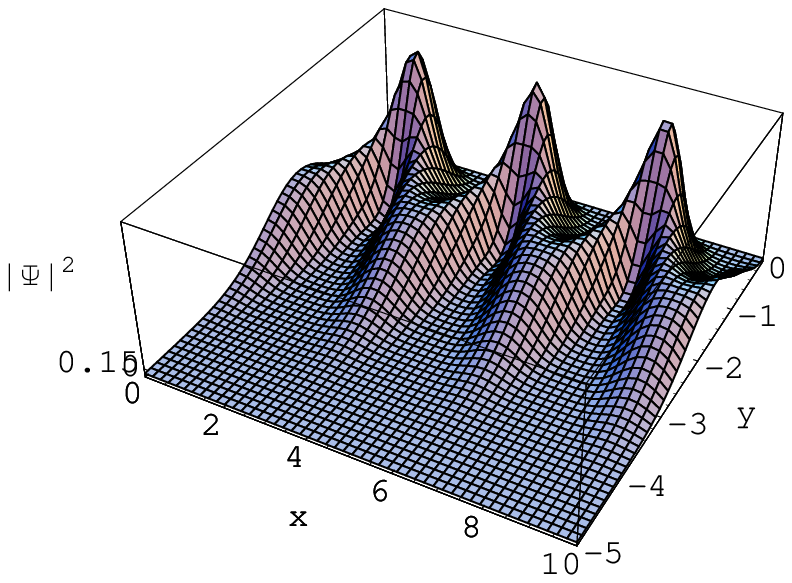,width=5cm}
\hspace{1.5cm} \epsfig{figure=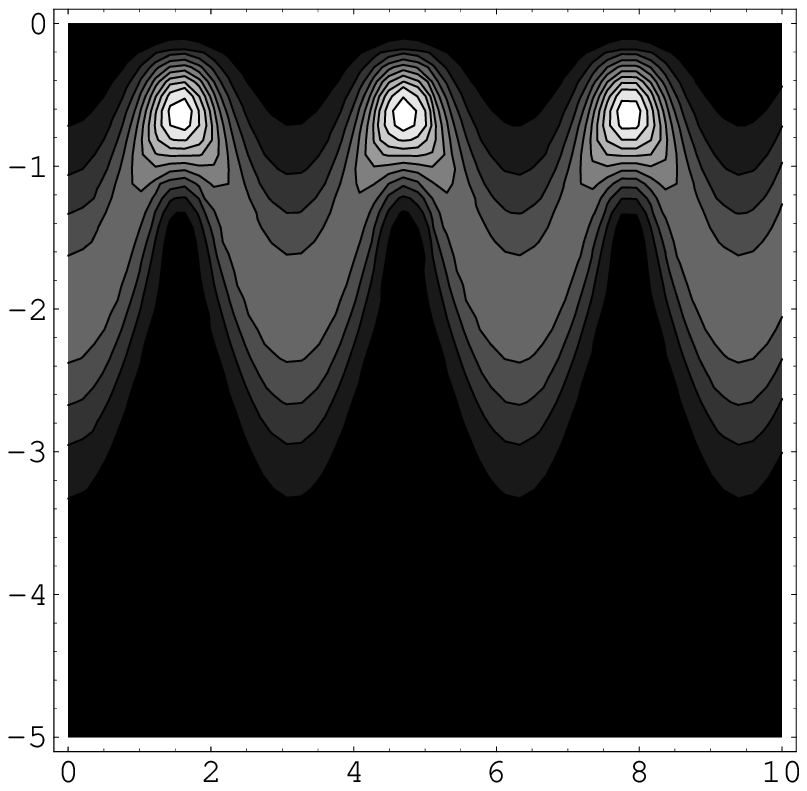,width=3.5cm}
\end{tabular}
\begin{tabular}{c}\epsfig{figure=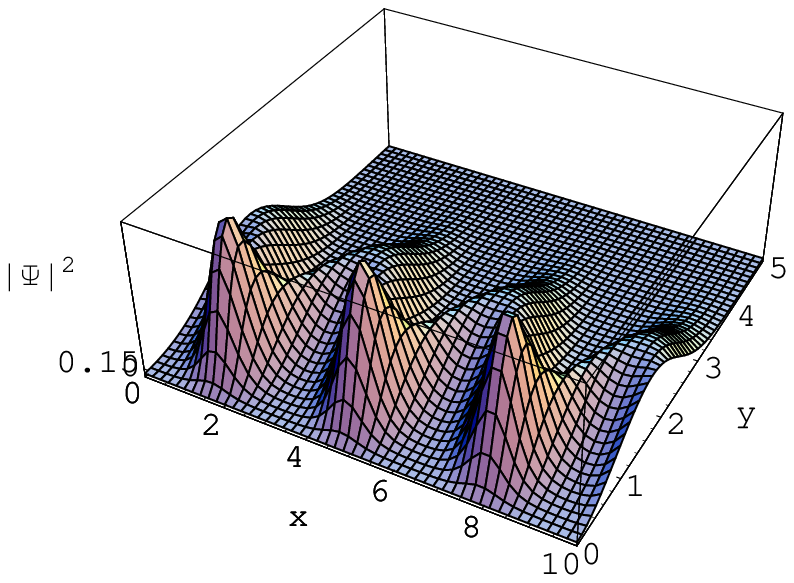,width=5cm}
\hspace{1.5cm} \epsfig{figure=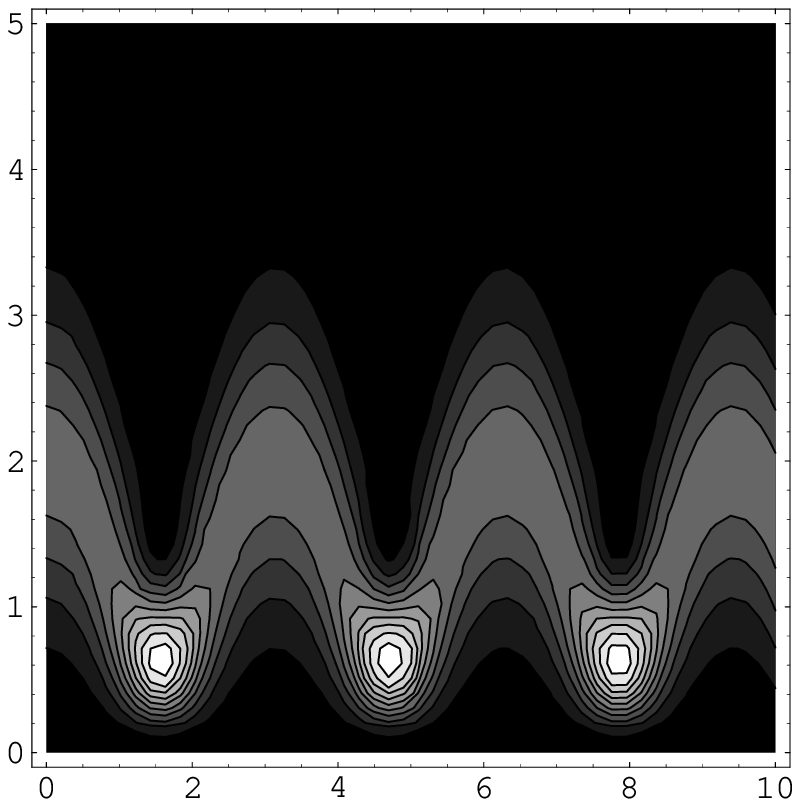,width=3.5cm}
\end{tabular}
\caption{\footnotesize  The figures show $|\Psi(X,Y)|^2$, the
square of the wave function and its corresponding contour plot.
Up: the figures are plotted for negative values of $y$ and bottom:
the same figures are plotted for positive values of $y$.
}\label{fig1}
\end{figure}

This solution is related to
the Noether symmetry generator $X=\frac{\partial}{\partial x}$ and
Noether conserved charge $Q=P_x$. In this case from equation
(\ref{AC}) we obtain
\begin{equation}\label{AF}
V\left(\frac{y}{x}\right)=\frac{y^2}{y^2-x^2}.\end{equation}In
terms of the old variables $(a, \phi)$ this potential is
$V(\phi)\sim \sinh^2(\omega \phi)$. With the choose of this
potential the Hamiltonian of the model becomes
\begin{equation}\label{AG}
{\cal H}=\frac{1}{2}\left(P_x^2-P_y^2\right)-\frac{1}{2}\omega^2
y^2.\end{equation}This Hamiltonian is nothing but the difference
of a free particle and a harmonic oscillator Hamiltonians.
Classically, this means that the motion in $y$-direction is
bounded and quantum mechanically it becomes natural to require the
corresponding wave function to be normalizable. Now, the classical
and quantum solutions of the model described by Hamiltonian
(\ref{AG}) can be easily obtained. The classical dynamics is
governed by the Hamiltonian equations, that is

\begin{eqnarray}\label{AH}
\left\{
\begin{array}{ll}
\dot{x}=\{x,{\cal H}\}=P_x,\hspace{0.5cm}\dot{P_x}=\{P_x,{\cal H}\}=0,\\\\
\dot{y}=\{y,{\cal H}\}=-P_y,\hspace{0.5cm}\dot{P_y}=\{P_y,{\cal H}\}=\omega^2 y,\\
\end{array}
\right.
\end{eqnarray}These equations can be immediately integrated to
yield
\begin{equation}\label{AI}
x(t)=P_{0x}t+x_0,\hspace{0.5cm}P_x=P_{0x},\end{equation} and
\begin{equation}\label{AJ}
y(t)=\frac{P_{0x}}{\omega}\sin \left(\omega
t+\delta_0\right),\hspace{0.5cm}P_y(t)=-P_{0x}\cos \left(\omega
t+\delta_0\right),\end{equation}where $P_{0x}$, $x_0$ and
$\delta_0$ are integration constants.  Since in the quantum
version of the model we are interested in constructing wave
packets from the WDW equation, we would like to obtain a classical
trajectory in configuration space $(x,y)$, where the classical
time $t$ is eliminated. This is because no such parameter exists
in the WDW equation. It is easy to see that the classical
solutions (\ref{AI}) and (\ref{AJ}) may be displayed as the
following trajectories
\begin{equation}\label{AK}
y=\frac{P_{0x}}{\omega}\sin \left[\frac{\omega}{P_{0x}}(x-x_0)
+\delta_0\right].\end{equation}Going back to the old variables
$(a,\phi)$ we get the cosmological solutions from (\ref{V}) as
\begin{equation}\label{AL}
a(t)=\omega^{2/3}\left[(P_{0x}t+x_0)^2-\frac{P_{0x}^2}{\omega^2}\sin^2(\omega
t+\delta_0)\right]^{1/3},\end{equation}and
\begin{equation}\label{AM}
\phi(t)=\frac{1}{\omega}\mbox{Arctanh}\left[\frac{\frac{P_{0x}}{\omega}\sin
\left(\omega
t+\delta_0\right)}{P_{0x}t+x_0}\right].\end{equation}We now focus
attention on the study of the quantization of the model described
above. This can be achieved via canonical quantization procedure
which leads the WDW equation. For the Hamiltonian (\ref{AG}) this
equation reads
\begin{equation}\label{AN}
\left(-\frac{\partial^2}{\partial x^2}+\frac{\partial^2}{\partial
y^2}-\omega^2 y^2\right)\Psi(x,y)=0.\end{equation}We separate the
variables in this equation as $\Psi(x,y)=e^{i\nu x}{\cal Y}(y)$
leading to $\frac{d^2 {\cal Y}}{dy^2}+(\nu^2-\omega^2 y^2){\cal
Y}(y)=0$, where $\nu$ is a separation constant. The physical
acceptable solutions to this equation can be found in terms of the
Hermite polynomials $H_n(z)$, if
$\nu^2=(2n+1)\omega=\nu_n^2\omega$, for some integer $n$. Thus,
the eigenfunctions of the WDW equation can be written as
\begin{equation}\label{AO}
\Psi_n(X,Y)=e^{i\nu_n
X}e^{-\frac{1}{2}Y^2}H_n(Y),\end{equation}where
$(X,Y)=\sqrt{\omega}(x,y)$. There still remains the question of
the boundary conditions on the solutions to the WDW equation. Note
that the minisuperspace of the above model is a two dimensional
manifold $0<a<\infty$, $-\infty<\phi<+\infty$ . According to
\cite{Vil}, its nonsingular boundary is the line $a=0$ with
$|\phi|<\infty$, while at the singular boundary, at least one of
the two variables is infinite. In terms of the variables $x$ and
$y$, introduced in (\ref{V}), the minisuperspace is recovered by
$x>0$, $x>|y|$, and the nonsingular boundary may be represented by
$x=y=0$. Since the minisuperspace variables are restricted to the
above mentioned domain, the minisuperspace quantization  deals
only with wave functions defined on this region.  Therefore, to
construct the quantum version of the model one should take into
account this issue. This is because that in such cases one usually
has to impose boundary conditions on the allowed wave functions
otherwise the relevant operators, specially the Hamiltonian, will
not be self-adjoint. The condition for the Hamiltonian operator
$\hat{{\cal H}}$ associated with the classical Hamiltonian
function (\ref{X}) to be self-adjoint is $(\psi_1,\hat{{\cal
H}}\psi_2)=(\hat{{\cal H}}\psi_1,\psi_2)$ or
\begin{eqnarray}\label{AO1}
\int_{\Omega} \psi_1^*(x,y)\hat{{\cal H}}\psi_2(x,y)dx dy=
\int_{\Omega} \psi_2(x,y)\hat{{\cal H}}\psi_1^*(x,y)dx
dy,\end{eqnarray}where the integrals should be taken over the
domain where the minisuperspace variables are defined on which.
Following the calculations in \cite{Lem} and dealing only with
square integrable wave functions, this condition yields a
vanishing wave function at  nonsingular boundary of the
minisuperspace. Hence, we impose the boundary condition on the
solutions (\ref{AO}) such that at the nonsingular boundary (at
$a=0$ and $|\phi|<\infty$) the wave function vanishes. This makes
the Hamiltonian hermitian and self-adjoint and can avoid the
singularities of the classical theory, i.e. there is zero
probability for observing a singularity corresponding to $a=0$.
Therefore, we require
\begin{equation}\label{AQ}
\Psi(a=0,\phi)=0\Rightarrow \Psi(X=0,Y=0)=0,\end{equation}which
yields
\begin{equation}\label{AP}
H_n(0)=0\Rightarrow n=\mbox{odd}.\end{equation} In general, one of
the most important features in quantum cosmology is the recovery
of classical cosmology from the corresponding quantum model or, in
other words, how can the WDW wave functions predict a classical
universe. In this approach, one usually constructs a coherent wave
packet with good asymptotic behavior in the minisuperspace,
peaking in the vicinity of the classical trajectory \cite{Mar}.
Therefore, we may now write the general solutions to the WDW
equation as a superposition of the eigenfunctions, that is
\begin{equation}\label{AR}
\Psi(X,Y)=\sum_{n=0}^\infty c_n e^{i\nu_n
X}e^{-\frac{1}{2}Y^2}H_{2n+1}(Y),\end{equation}where $c_n$ are
suitable weight factors to construct the wave packets and to make
a convergent superposition, we may choose them such that the
summands with smaller $n$ have more important contribution to the
above superposition. Therefore, to achieve an analytical
expression for the wave function we assume that the above
superposition is taken over such values of $n$ for which one can
use the approximation $\nu_n \sim 2n+1$, so that
\begin{equation}\label{AS}
\Psi(X,Y)=\sum_{n=0}^\infty c_n e^{i(2n+1)
X}e^{-\frac{1}{2}Y^2}H_{2n+1}(Y).\end{equation}By using the
equality
\begin{equation}\label{AT}
\sum_{n=0}^\infty
\frac{\gamma^{2n+1}}{(2n+1)!}H_{2n+1}(z)=e^{-\gamma^2}\sinh2\gamma
z,\end{equation}we can evaluate the sum over $n$ in (\ref{AS}) and
simple analytical expression for this sum is found if we choose
$\gamma$ to be $e^{iX}$ and $c_n={\cal N}/(2n+1)!$, which results
in\footnote{One may have some doubts on this final form for the
wave function and the following results due to the assumption
$\nu_n \sim 2n+1$ which seems to be irrelevant especially in the
case of a discrete spectrum for $n$. To overcome this problem, we
have made a numerical study of the behavior of $|\Psi|^2$ based on
equation (\ref{AR}) with $c_n=1/(2n+1)!$ and $\nu_n=(2n+1)^2$ and
have verified that the general patterns of the resulting wave
packets follow the behavior shown in figure \ref{fig1}  with a
very good approximation.}
\begin{equation}\label{AU}
\Psi(X,Y)={\cal
N}e^{-\frac{1}{2}Y^2}\exp(-e^{2iX})\sinh\left(2Ye^{iX}\right),\end{equation}where
${\cal N}$ is a numerical factor. Before going any further, some
remarks are in order. An important ingredient in any model theory
related to the quantization of a cosmological setting is the
choice of a probability measure in order to make predictions from
a given solution to the WDW equation. Since the WDW equation is a
second order Klaein-Gordon type equation in quantum field theory,
the most widely used probability density for it results the
conserved current $j=\frac{i}{2}(\Psi^*\nabla \Psi-\Psi \nabla
\Psi^*)$. But this immediately presents a problem, because the
probability density given by this expression is not positive
definite. For this reason and some other difficulties related to
this interpretation, an alternative used approach is to introduce
$|\Psi|^2$ as the probability density. Although this definition is
necessarily positive definite and specially works very well for
the minisuperspace models, it suffers from some problems. The most
important of such problems arise when we are dealing with the
non-normalizable. Despite of the existence of the above mentioned
problems, here we shall look for the peaks in the wave function in
the sense that if a peak is sufficiently strong in some region,
the probability of the universe being in this region is large. In
figure \ref{fig1} we have plotted the square of the wave function
(\ref{AU}) and its corresponding contour plot. It is seen that the
peaks of the wave function follow a periodic pattern. In
comparison with the classical path (\ref{AK}), this means that a
good correlation exists between the quantum patterns shown in this
figure and the classical trajectory (\ref{AK}) in configuration
space $(x,y)$.

\subsubsection{The case: $(\alpha,\beta)=(0,1)$}This case relates
the Noether vector $\frac{\partial}{\partial y}$ and its
corresponding conserved charge is $Q=P_y$. From (\ref{AC}) the
potential function reads as
\begin{equation}\label{AV}
V\left(\frac{y}{x}\right)=\frac{x^2}{x^2-y^2},\end{equation}where
in terms of the old variables $(a,\phi)$ can be termed as
$V(\phi)\sim \cosh^2\left(\omega \phi\right)$. Since the
Hamiltonian of the model, in this case is
\begin{equation}\label{AW}
{\cal H}=\frac{1}{2}\left(P_x^2-P_y^2\right)+\frac{1}{2}\omega^2
x^2,\end{equation}we see that there is no major difference between
this case the the previous one except that the variables $x$ and
$y$ change their role with each other. Therefore, following the
steps (\ref{AG})-(\ref{AU}), we are led to the classical solutions
\begin{equation}\label{AX}
x(t)=\frac{P_{0y}}{\omega}\sin\left(\omega
t+\eta_0\right),\hspace{0.5cm}P_x=P_{0y}\cos\left(\omega
t+\eta_0\right),\end{equation}

\begin{equation}\label{AY}
y(t)=-P_{0y}t+y_0,\hspace{0.5cm}P_y(t)=P_{0y},\end{equation}and
the corresponding classical trajectories
\begin{equation}\label{AZ}
x=\frac{P_{0y}}{\omega}\sin\left[\frac{\omega}{P_{0y}}
(y_0-y)+\eta_0\right],\end{equation}where in terms of the old
variables $(a,\phi)$ can be viewed as

\begin{equation}\label{BA}
a(t)=\omega^{2/3}\left[\frac{P_{0y}^2}{\omega^2}\sin^2(\omega
t+\eta_0)-(-P_{0y}t+y_0)^2\right]^{1/3},\end{equation}and

\begin{equation}\label{BC}
\phi(t)=\frac{1}{\omega}\mbox{Arctanh}\left[\frac{-P_{0y}t+y_0}{\frac{P_{0y}}{\omega}\sin\left(\omega
t+\eta_0\right)}\right].\end{equation}Also, the wave function of
the quantum counterpart of the model can be written as
\begin{equation}\label{BD}
\Psi(X,Y)={\cal
N}e^{-\frac{1}{2}X^2}\exp(-e^{2iY})\sinh\left(2Xe^{iY}\right).\end{equation}The
discussions on the comparison between quantum cosmological
solution and its classical version are the same as previous model,
namely the $(\alpha,\beta)=(1,0)$ model. Similar discussion as
above would be applicable to this case as well.

\subsubsection{The case: $(\alpha,\beta)=(y,x)$}In this case the
Noether symmetry is generated by the following vector field
\begin{equation}\label{BE}
X=y\frac{\partial}{\partial x}+x\frac{\partial}{\partial
y}+\dot{y}\frac{\partial}{\partial
\dot{x}}+\dot{x}\frac{\partial}{\partial
\dot{y}},\end{equation}which corresponds to the Noether conserved
charged $Q=yP_x+xP_y$. The condition (\ref{AC}), in this case,
demands that the potential function be a constant function and as
such, we choose it to be $1$. Therefore, the Hamiltonian takes the
form
\begin{equation}\label{BF}
{\cal
H}=\frac{1}{2}\left(P_x^2-P_y^2\right)+\frac{1}{2}\omega^2\left(x^2-y^2\right),\end{equation}which
describes an isotropic oscillator-ghost-oscillator system. The
classical equations of motion are given by
\begin{eqnarray}\label{BG}
\left\{
\begin{array}{ll}
\dot{x}=\{x,{\cal H}\}=P_x,\hspace{0.5cm}\dot{P_x}=\{P_x,{\cal H}\}=-\omega^2 x,\\\\
\dot{y}=\{y,{\cal H}\}=-P_y,\hspace{0.5cm}\dot{P_y}=\{P_y,{\cal H}\}=\omega^2 y.\\
\end{array}
\right.
\end{eqnarray}Choosing the integration constants such that the solutions satisfy the zero energy condition ${\cal H}=0$, the
solutions are obtained as
\begin{eqnarray}\label{BH}
\left\{
\begin{array}{ll}
x(t)=A\sin\left(\omega t+\eta_0\right),\hspace{0.5cm}P_x(t)=A\omega\cos\left(\omega t+\eta_0\right),\\\\
y(t)=\ell A \sin\left(\omega t+\delta_0\right)\hspace{0.5cm}P_y(t)=-\ell A \omega \cos\left(\omega t+\delta_0\right),\\
\end{array}
\right.
\end{eqnarray}where $A$, $\delta_0$ and $\eta_0$ are integration
constants and $\ell=\pm 1$. From the above equations, we see that
the classical trajectories obey the relation
\begin{equation}\label{BI}
y^2+x^2-2\ell xy
\cos(\delta_0-\eta_0)-A^2\sin^2(\delta_0-\eta_0)=0.\end{equation}This
equation describes ellipses which their major axes make angle
$\pi/4$ with the positive/negative $x$-axis according to the
choices $\pm 1$ for $\ell$. Also, the eccentricity and the size of
each trajectory are determined by $(\delta_0-\eta_0)$ and $A$
respectively. Now, using the transformation (\ref{V}) the
classical cosmological behavior can be obtained as
\begin{equation}\label{BJ}
a(t)=A^{2/3}\omega^{2/3}\sin^{1/3}(\delta_0-\eta_0)\sin^{1/3}\left(2\omega
t+\delta_0+\eta_0\right),\end{equation}and
\begin{equation}\label{BK}
\phi(t)=\frac{1}{\omega}\mbox{Arctanh}\left[\frac{\ell
\sin\left(\omega t+\delta_0\right)}{\sin\left(\omega
t+\eta_0\right)}\right].\end{equation}At this step, as the
previous subsections, we deal with the quantization of the model.
The WDW equation corresponding to the Hamiltonian (\ref{BF}) reads
\begin{equation}\label{BL}
\left(-\frac{\partial^2}{\partial x^2}+\frac{\partial^2}{\partial
y^2}+\omega^2 x^2-\omega^2 y^2\right)\Psi(x,y)=0.\end{equation}
This equation is a quantum isotropic oscillator-ghost-oscillator
system with zero energy. Therefore, its solutions belong to a
subspace of the Hilbert space spanned by separable eigenfunctions
of a two-dimensional isotropic simple harmonic oscillator
Hamiltonian. In \cite{Car} the models of coupled harmonic
oscillators have been studied in other contexts relevant for
quantum cosmology. Separating the eigenfunctions of (\ref{BL}) in
the form
\begin{equation}\label{BM}
\Psi_{n_1,n_2}(X,Y)={\cal X}_{n_1}(X){\cal Y}_{n_2}(Y),
\end{equation}
yields
\begin{equation}\label{BN}
{\cal
X}_{n_1}(X)=e^{-\frac{1}{2}X^2}H_{n_1}(X),\hspace{0.5cm}{\cal
Y}_{n_2}(Y)=e^{-\frac{1}{2}Y^2}H_{n_2}(Y),\end{equation}subject to
the restriction $n_1= n_2 = n$. In (\ref{BN}), $H_n(z)$ are the
Hermite polynomials and $(X,Y)=\sqrt{\omega}(x,y)$ as before.
Applying again the boundary condition (\ref{AQ}), we are led to
the following general solution to the WDW equation
\begin{equation}\label{BO}
\Psi(X,Y)=\sum_{n_1=0}^{\infty}\sum_{n_2=0}^{\infty}c_{n_1}c_{n_2}e^{-\frac{1}{2}(X^2+Y^2)}H_{2n_1+1}(X)H_{2n_2+1}(Y).\end{equation}
If we choose the coefficients $c_{n_1}$ and $c_{n_2}$ to be
$c_{n_1}=\xi^{2n_1+1}/(2n_1+1)!$ and
$c_{n_2}=\zeta^{2n_2+1}/(2n_2+1)!$ where $\xi$ and $\zeta$ are
arbitrary complex constants, the wave function takes the form
\begin{equation}\label{BP}
\Psi(X,Y)={\cal
N}e^{-\frac{1}{2}(X^2+Y^2)}e^{-(\xi^2+\zeta^2)}\sinh\left(2\xi
X\right)\sinh\left(2\zeta Y\right).\end{equation}

\begin{figure}
\begin{tabular}{c}\epsfig{figure=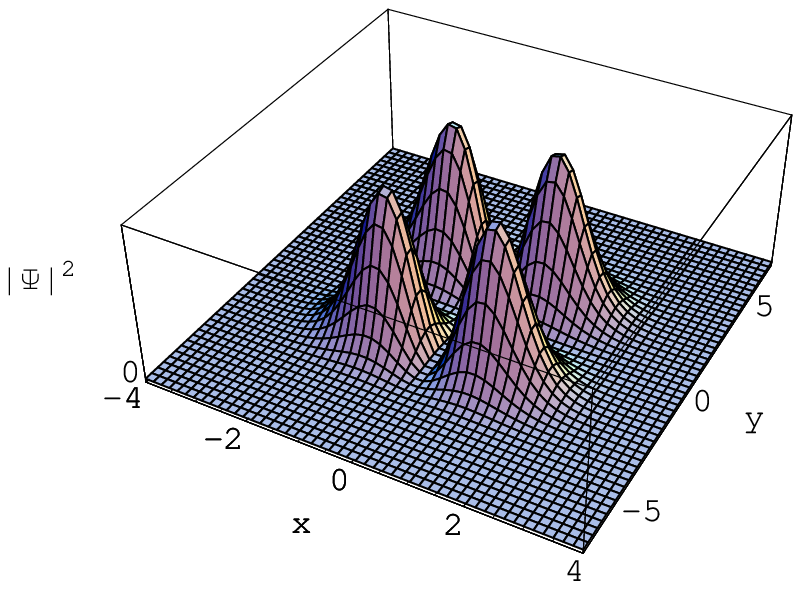,width=5cm}
\hspace{1.5cm} \epsfig{figure=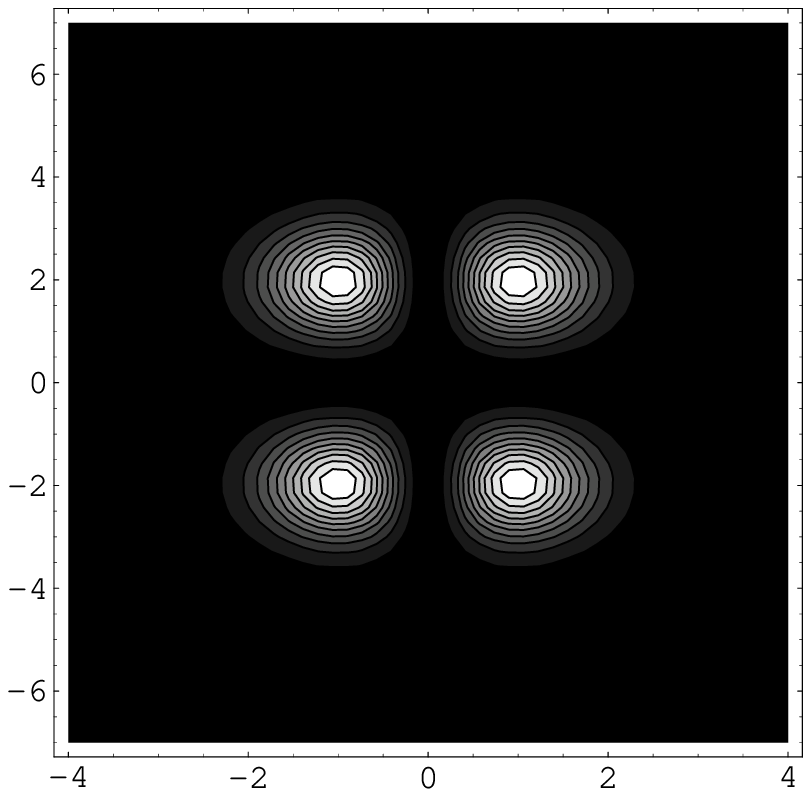,width=3.5cm}\hspace{1.5cm}
\epsfig{figure=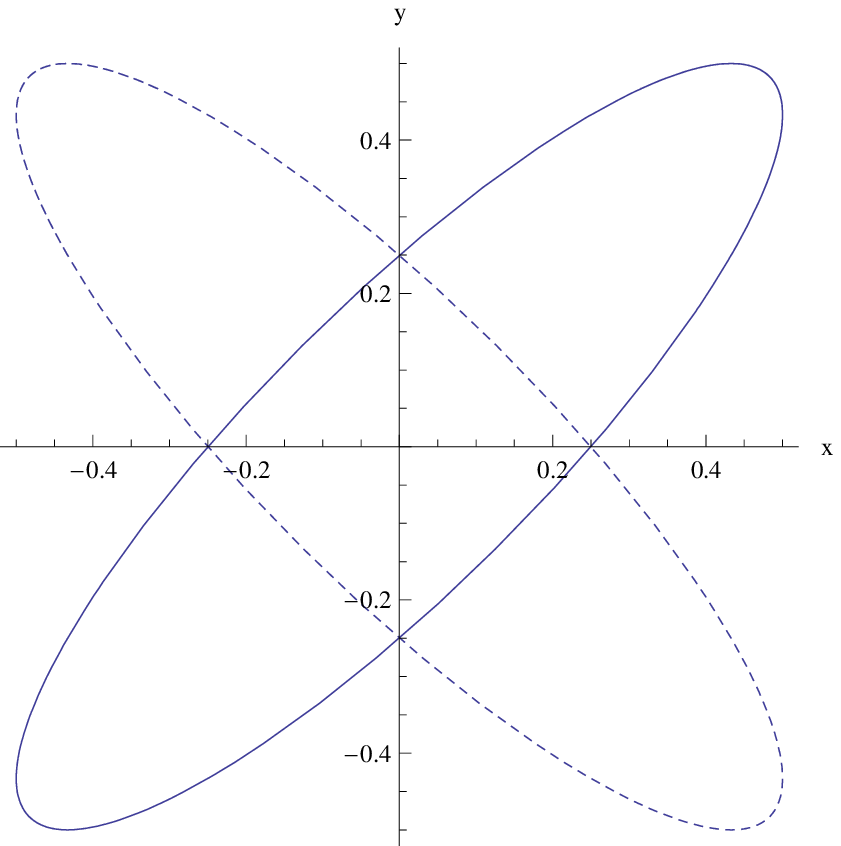,width=3.5cm}
\end{tabular}
\caption{\footnotesize  The figures show $|\Psi(X,Y)|^2$, the
square of the wave function (\ref{BP}), its corresponding contour
plot and the classical trajectory (\ref{BI}), (solid line for
$\ell=1$ and dashed line for $\ell=-1$).}\label{fig2}
\end{figure}
In figure \ref{fig2}, we have shown the qualitative behavior of
this wave function and its contour plot. To realize the
correlation between these quantum patterns and the classical
trajectories represented by (\ref{BI}), note that in the
minisuperspace formulation, the cosmic evolution of the universe
is modeled with the motion of a point particle in a space with
minisuperspace coordinates. In motion on an ellipse, the particle
(universe) spend most of its time in the region near its apogees.
This means that the particle will be found around the apogees of
its trajectory with the maximum probability. This is just what the
figure \ref{fig2} is showing. As it is seen from this figure, the
wave function has two pair peaks, each pair may be interpreted as
two apogees of the classical trajectory. In classical model, the
transition from one point on the configuration space to any other
points will be described by a continuous motion on the classical
trajectory. The quantum description of such a transition, on the
other hand, may be explained by a tunneling procedure. This means
that there are different possible states from which our present
universe could have evolved and tunneled in the past, from one
state to another. Again, we see that our quantization leads us to
a model free of the classical singularity and a good correlation
with its classical counterpart.

\subsection{Phantom scalar field}
In this case ($\epsilon=-1$), the minisuperspace manifold would
have an Euclidean signature (see (\ref{U})) with zero Ricci
scalar. To write the metric in its canonical form, we apply the
transformation

\begin{equation}\label{BR}
x=\frac{1}{\omega}a^{3/2}\cos (\omega
\phi),\hspace{0.5cm}y=\frac{1}{\omega}a^{3/2}\sin (\omega
\phi).\end{equation} It is easy to see that in terms of these new
variables the metric takes the form of a $2$-dimensional Euclidean
one represented by

\begin{equation}\label{BS}
\bar{{\cal G}}_{AB}dQ^AdQ^B=-dx^2-dy^2.\end{equation}Also, the
phantom counterpart of Lagrangian (\ref{T}) becomes (we consider
$k=0$)

\begin{equation}\label{BT}
{\cal
L}=\frac{1}{2}\left(-\dot{x}^2-\dot{y}^2\right)-\frac{1}{2}\omega^2\left(x^2+y^2\right)
V\left(\frac{y}{x}\right),\end{equation}with the corresponding
Hamiltonian becoming
\begin{equation}\label{BU}
{\cal
H}=\frac{1}{2}\left(-P_x^2-P_y^2\right)+\frac{1}{2}\omega^2\left(x^2+y^2\right)V\left(\frac{y}{x}\right).\end{equation}Again,
considering a Noether symmetry generated by a vector field like
(\ref{Z}), the condition $L_X {\cal L}=0$ will lead us to the
following equations for $\alpha$ and $\beta$
\begin{equation}\label{BV}
-\omega^2V(z)(\beta y+\alpha x)+\frac{1}{2}\omega^2
V'(z)\left(x^2+y^2\right)\left(\alpha \frac{y}{x^2}-\beta
\frac{1}{x}\right)=0,\end{equation}
\begin{equation}\label{BW}
\frac{\partial \alpha}{\partial x}=\frac{\partial \beta}{\partial
y}=0,\hspace{0.5cm}\frac{\partial \alpha}{\partial
y}+\frac{\partial \beta}{\partial x}=0,\end{equation}where $z=y/x$
and $V'(z)=dV/dz$ as before. The independent solutions of the
system (\ref{BW}) my be obtained as

\begin{equation}\label{BX}
(\alpha,\beta)=(1,0),\hspace{0.5cm}(\alpha,\beta)=(0,1),\hspace{0.5cm}(\alpha,\beta)=(y,-x).\end{equation}Now,
following the same procedure as in the previous subsection, we
consider the above tree case separately to study their
corresponding classical and quantum cosmology.
\subsubsection{The case: $(\alpha, \beta)=(1,0)$}
\begin{figure}
\begin{tabular}{c}\epsfig{figure=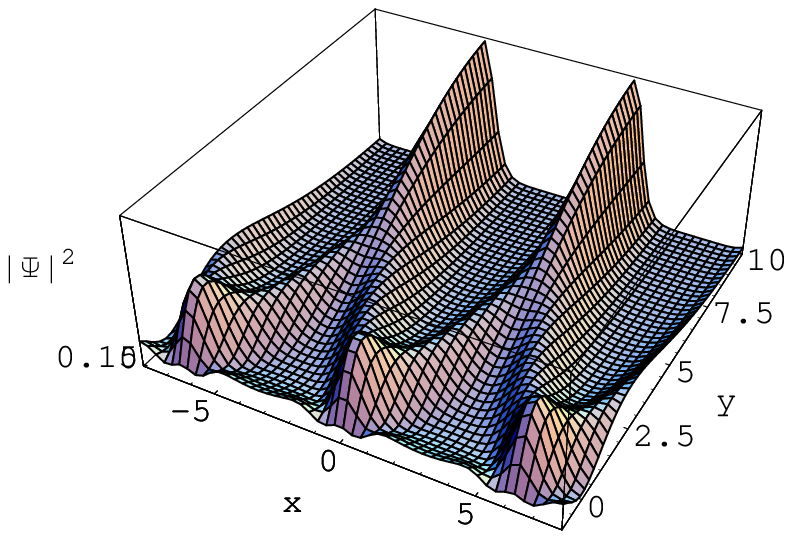,width=5cm}
\hspace{1.5cm} \epsfig{figure=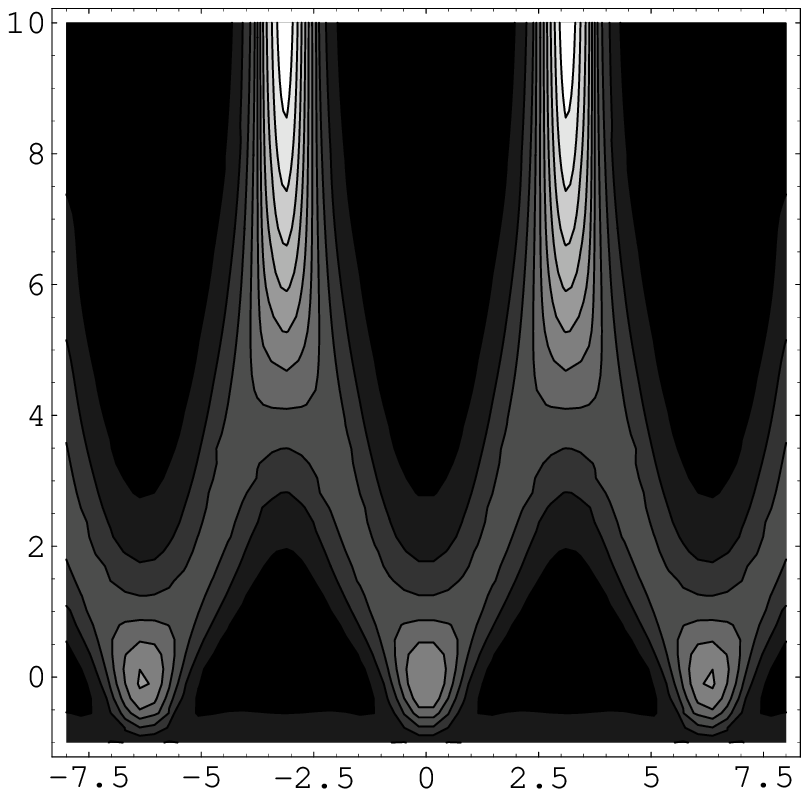,width=3.5cm}\hspace{1.5cm}
\epsfig{figure=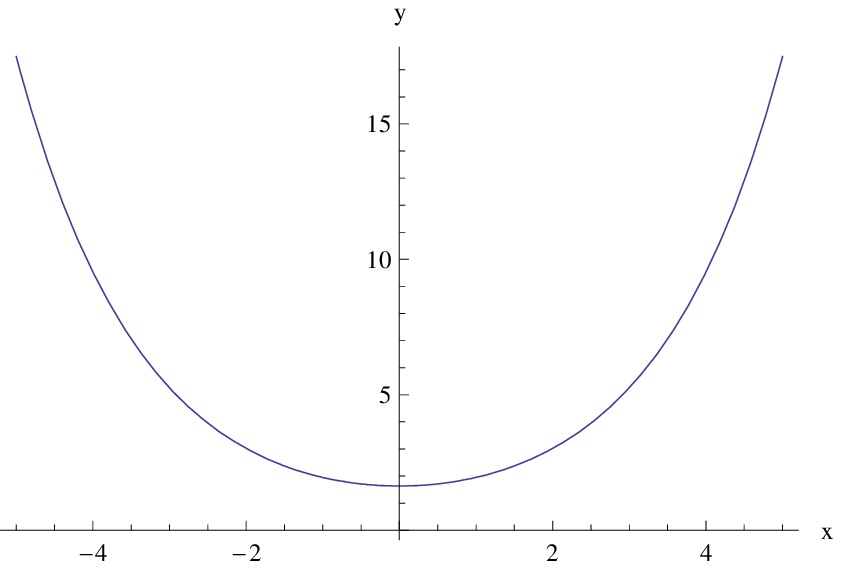,width=4cm}
\end{tabular}
\caption{\footnotesize  From left to right: $|\Psi(X,Y)|^2$, the
square of the phantom wave function (\ref{CI}), in which we have
taken the combinations from six terms (taking more terms would
only have minor effects on the shape of the wave function) with
$c_n=1/(n! 2^n)$, its corresponding contour plot and the classical
trajectory (\ref{CE}).}\label{fig3}
\end{figure}

This case corresponds to the vector field
$X=\frac{\partial}{\partial x}$ which results the conserved
quantity $Q=P_x$. The potential function can be obtained from
(\ref{BV}) as

\begin{equation}\label{BY}
V\left(\frac{y}{x}\right)=\frac{y^2}{x^2+y^2},\end{equation}where
in terms of the cosmic variables $(a, \phi)$ may be written as
$V(\phi)\sim \sin^2\left(\omega \phi\right)$. This potential
function transforms the Hamiltonian (\ref{BU}) into the form
\begin{equation}\label{BZ}
{\cal H}=\frac{1}{2}\left(-P_x^2-P_y^2\right)+\frac{1}{2}\omega^2
y^2,\end{equation}from which one gets the following classical
equations of motion
\begin{eqnarray}\label{CA}
\left\{
\begin{array}{ll}
\dot{x}=\{x,{\cal H}\}=-P_x,\hspace{0.5cm}\dot{P_x}=\{P_x,{\cal H}\}=0,\\\\
\dot{y}=\{y,{\cal H}\}=-P_y,\hspace{0.5cm}\dot{P_y}=\{P_y,{\cal H}\}=-\omega^2 y,\\
\end{array}
\right.
\end{eqnarray}with solutions
\begin{equation}\label{CB}
x(t)=-P_{0x}t+x_0,\hspace{0.5cm}P_x=P_{0x},\end{equation} and
\begin{equation}\label{CD}
y(t)=\frac{P_{0x}}{\omega}\cosh \left(\omega
t+\delta_0\right),\hspace{0.5cm}P_y(t)=-P_{0x}\sinh \left(\omega
t+\delta_0\right),\end{equation}where for the sake of the
simplicity, we use the same notation as in the previous section
for the integration constants. These solutions obey the classical
trajectories in the configuration space $(x,y)$
\begin{equation}\label{CE}
y=\frac{P_{0x}}{\omega}\cosh
\left(\frac{\omega}{P_{0x}}(x_0-x)+\delta_0\right).\end{equation}Finally,
going back to the original variables $(a,\phi)$, the corresponding
classical cosmology will be obtained as
\begin{equation}\label{CF}
a(t)=\omega^{2/3}\left[(-P_{0x}t+x_0)^2+\frac{P_{0x}^2}{\omega^2}\cosh^2(\omega
t+\delta_0)\right]^{1/3},\end{equation}and
\begin{equation}\label{CG}
\phi(t)=\frac{1}{\omega}\mbox{Arctan}\left[\frac{\frac{P_{0x}}{\omega}\cosh
\left(\omega
t+\delta_0\right)}{-P_{0x}t+x_0}\right].\end{equation}We see that
unlike the case of the ordinary scalar field, here all the motions
are unbounded, so that it becomes natural to expect that the wave
function describing the quantum version of the model to be
non-normalizable.  Now, by canonical quantization the WDW equation
is given by

\begin{equation}\label{CH}
\left(\frac{\partial^2}{\partial x^2}+\frac{\partial^2}{\partial
y^2}+\omega^2 y^2\right)\Psi(x,y)=0.\end{equation}The
eigenfunctions of this equation can be found by the method of
separation of variables, by means of which we obtain
\begin{equation}\label{CH1}
\Psi_{\nu}(X,Y)=e^{\pm i\nu X}{\cal D}_{-\frac{1}{2}\pm
i\frac{\nu^2}{2}}\left((\pm 1+i)Y\right),\end{equation}where $\nu$
is the separation constant and ${\cal D}$ is the parabolic
cylinder function. As before the general solutions may be
constructed by a superposition of these eigenfunctions. Since
non-normalizable states in this case belong to the continuous
spectrum, the general wave function should be constructed by a
continuous superposition in the form
\begin{equation}\label{CH2}
\Psi(X,Y)=\int_{-\infty}^{+\infty}C(\nu)\Psi_{\nu}(X,Y)d\nu.\end{equation}
However, to achieve a more clear picture of the wave function by
the numerical methods which would be comparable with the results
in the previous section, we restrict ourselves to a discrete
superposition of the above eigenfunctions. Thus, by choosing the
separation constant as $\nu^2=(2n-i)\omega=\nu_n^2 \omega$ and
connecting the parabolic cylinder function with the Hermite
functions \cite{Ab}, the general solution to the WDW equation
takes the form
\begin{equation}\label{CI}
\Psi(X,Y)=\sum_{n=0}^\infty c_n e^{\pm i\nu_n
X}e^{-\frac{1}{2}iY^2}H_{in}\left(\frac{1}{\sqrt{2}}(1-i)Y\right).\end{equation}We
have summarized the above results in figure \ref{fig3}. As this
figure shows, like in the case of the ordinary scalar field, the
evolution of the universe based on the classical cosmology
represents a late time expansion. This is because the particle
(universe) moves on one of the branches of its classical
trajectory forever without any turning points. A remarkable point
about this motion is that in spite of the ordinary scalar field
case, it does not come from a big-bang singularity in which the
scale factor goes to zero, but instead tends to another kind of
singularity in which the scale factor diverges. In the models
where the phantom fields are considered by a perfect fluid with
the equation of state parameter less than $-1$, this kind of
singularity will be achieved in a finite-time and is called the
big-rip singularity. On the other hand a glance at the quantum
patterns shows that although a good correlation exists between
them and the classical loci in the configuration space, but the
quantum effects dominate in the region of the classical
singularity, i.e., at the large values of scale factor. In this
regime the quantum solutions fall from an expansion phase to a
contraction era and this phenomenon will be repeated cyclic. By
such a behavior we see that the quantum effects show their
important role for the large values of the scale factor and thus
the evolution of the scale factor towards a big-rip like
singularity will be avoided \cite{Bab}. However, a remark about
the above analysis is that it happens in some internal time
parameter such as $x$ and in order to see whether it is physically
meaningful the results should translate in terms of the cosmic
time $t$. The answer to this question in general may be not an
easy task. This is because one usually uses non-standard
parametrization of the metric in order to simplify the
calculations and have a manageable Lagrangian. Therefore,
returning the results into the proper time gauge is not always an
integrable process. By the way, in our simple model at hand, since
(\ref{CB}) represents a monotonic relation between $x$ and $t$, we
may argue that the above conclusions are also valid in terms of
the cosmic time gauge $t$ as well.

\subsubsection{The case: $(\alpha,\beta)=(0,1)$}The situation in
this case is alike the one that was occurred when we were
discussing about the ordinary scalar field. The potential function
resulting from this solution is $V(y/x)=x^2/(x^2+y^2)$, or
$V(\phi)\sim \cos^2(\omega \phi)$ in terms of the variables
$(a,\phi)$. Again, all of the results coming from this potential
function can be obtained from their counterparts in the previous
subsection with exchange the role of the variables $x$ and $y$
with each other and because of this we do not repeat the
calculations ones again.

\subsubsection{The case: $(\alpha,\beta)=(y,-x)$}
\begin{figure}
\begin{tabular}{c}\epsfig{figure=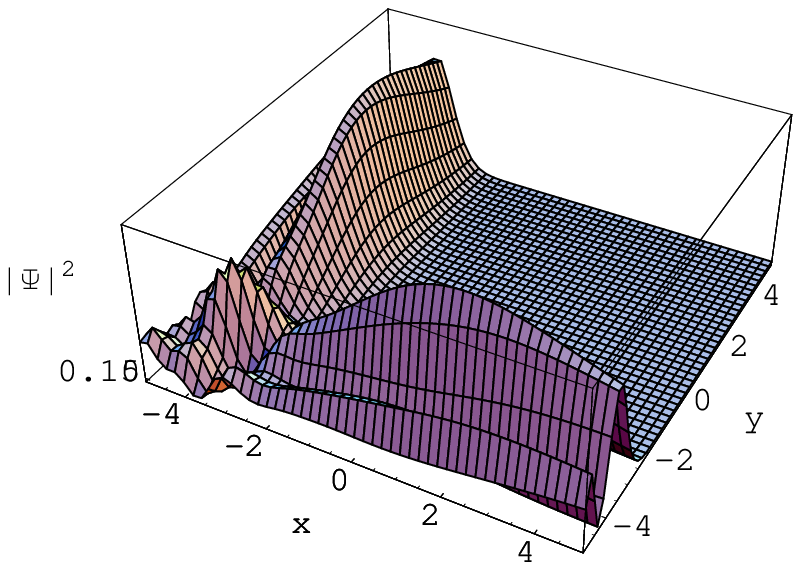,width=5cm}
\hspace{1.5cm} \epsfig{figure=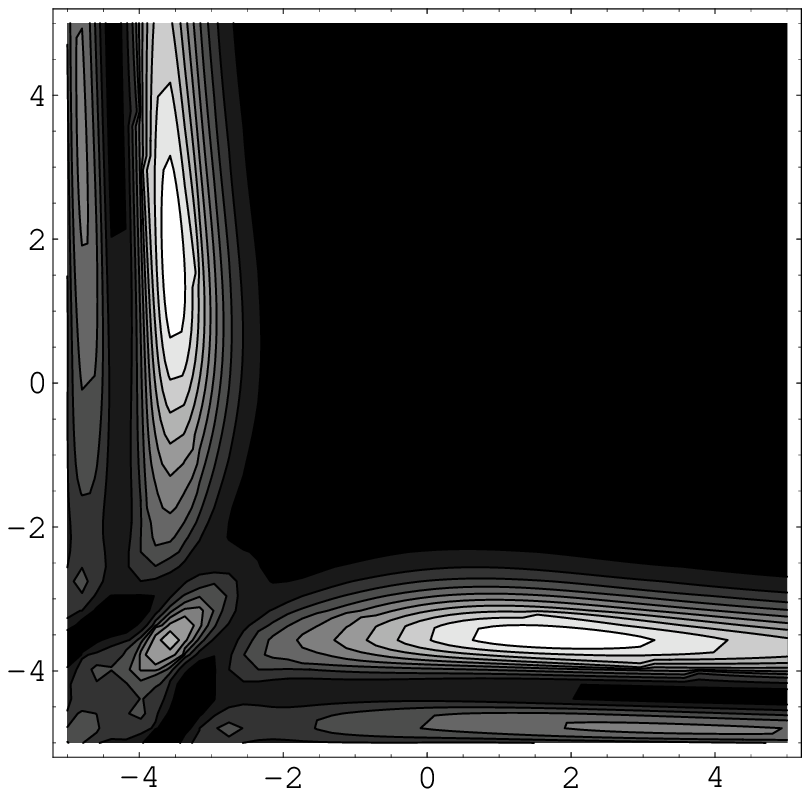,width=3.5cm}\hspace{1.5cm}
\epsfig{figure=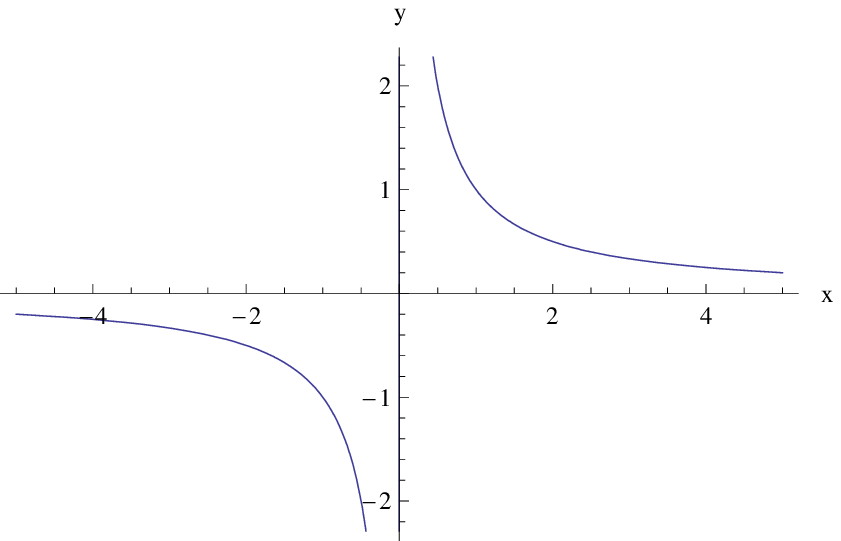,width=4cm}
\end{tabular}
\caption{\footnotesize  From left to right: $|\Psi(X,Y)|^2$, the
square of the phantom wave function (\ref{CR}), in which we have
taken the combinations from six terms (taking more terms would
only have minor effects on the shape of the wave function) with
$c_n=1/(n! 2^n)$, its corresponding contour plot and the classical
trajectory (\ref{CN}) with $\ell=1$.}\label{fig4}
\end{figure}

This solution generates the Noether symmetry with the vector field

\begin{equation}\label{CJ}
X=y\frac{\partial}{\partial x}-x\frac{\partial}{\partial
y}+\dot{y}\frac{\partial}{\partial
\dot{x}}-\dot{x}\frac{\partial}{\partial
\dot{y}},\end{equation}which gives the conserved charge
$Q=yP_x-xP_y$. Substitution of this solution into the equation
(\ref{BV}) we get a constant value for the potential function and
as before we take it to be equal to $1$. Therefore, the
corresponding dynamical system can be described by the Hamiltonian

\begin{equation}\label{CK}
{\cal
H}=\frac{1}{2}\left(-P_x^2-P_y^2\right)+\frac{1}{2}\omega^2\left(x^2+y^2\right).\end{equation}Following
the same steps as in the above sections, we arrive the classical
equations of motion as

\begin{eqnarray}\label{CL}
\left\{
\begin{array}{ll}
\dot{x}=\{x,{\cal H}\}=-P_x,\hspace{0.5cm}\dot{P_x}=\{P_x,{\cal H}\}=-\omega^2 x,\\\\
\dot{y}=\{y,{\cal H}\}=-P_y,\hspace{0.5cm}\dot{P_y}=\{P_y,{\cal H}\}=-\omega^2 y,\\
\end{array}
\right.
\end{eqnarray}which admit the following integrals

\begin{eqnarray}\label{CM}
\left\{
\begin{array}{ll}
x(t)=A\cosh\left(\omega t+\eta_0\right),\hspace{0.5cm}P_x(t)=-A\omega\sinh\left(\omega t+\eta_0\right),\\\\
y(t)=\ell A \sinh\left(\omega t+\delta_0\right)\hspace{0.5cm}P_y(t)=-\ell A \omega \cosh\left(\omega t+\delta_0\right),\\
\end{array}
\right.
\end{eqnarray}where as before we take $A$, $\delta_0$ and $\eta_0$ as integration
constants and $\ell=\pm 1$. Now one can eliminate the time
parameter $t$ from these solutions to get the classical
trajectories in the configuration space $(x,y)$ as
 \begin{equation}\label{CN}
 y^2-x^2-2\ell xy
 \sinh(\delta_0-\eta_0)+A^2\cosh^2(\delta_0-\eta_0)=0,\end{equation}which
 represents a hyperbola so that the the
particle (universe) has an unbounded motion in the $x-y$ plane.
Now, going back to the original variables $(a,\phi)$ with the help
of the transformation (\ref{BR}), the
 dynamics of the system reads as

\begin{equation}\label{CO}
a(t)=A^{2/3}\omega^{2/3}\cosh^{1/3}(\delta_0-\eta_0)\cosh^{1/3}\left(2\omega
t+\delta_0+\eta_0\right),\end{equation}and

\begin{equation}\label{CP}
\phi(t)=\frac{1}{\omega}\mbox{Arctan}\left[\frac{\ell
\sinh\left(\omega t+\delta_0\right)}{\cosh\left(\omega
t+\eta_0\right)}\right].\end{equation}To proceed, let us to deal
with the quantum cosmology associated with the model described by
the Hamiltonian (\ref{CK}). The corresponding WDW equation is

\begin{equation}\label{CQ}
\left(\frac{\partial^2}{\partial x^2}+\frac{\partial^2}{\partial
y^2}+\omega^2 x^2+\omega^2
y^2\right)\Psi(x,y)=0.\end{equation}With the same techniques as in
the previous sections and also with same notations one may
represent the general solutions to the above equation as

\begin{eqnarray}\label{CR}
\Psi(X,Y)=\sum_{n=0}^\infty c_n
e^{-\frac{1}{2}i(X^2+Y^2)}\left[H_{in}\left(\frac{1}{\sqrt{2}}(1-i)X\right)H_{-1-in}\left(\frac{1}{\sqrt{2}}(1-i)Y\right)+\nonumber
\right.\\ \left.
H_{in}\left(\frac{1}{\sqrt{2}}(1-i)Y\right)H_{-1-in}\left(\frac{1}{\sqrt{2}}(1-i)X\right)\right].\end{eqnarray}Figure
\ref{fig4} shows the qualitative behavior of the wave function and
classical trajectory for typical values of the parameters. As we
have mentioned above, the motion in configuration space is an
unbounded motion along one of the branches of a hyperbola. A point
(universe) with such a motion spend lots of its time far from the
apex of its path. This is exactly what we realize from the quantum
wave function in figure \ref{fig4}. These patterns show that the
peaks of the wave function follow the classical trajectory with a
good degree of accuracy.

\section{Conclusions}In this paper we have studied the scalar
field classical and quantum cosmology in a Noether symmetry point
of view. The minisuperspace of such a model is constructed of a
two dimensional manifold with vanishing Ricci scalar. Therefore,
it is possible to find a coordinate transformation which cast the
corresponding metric to a Minkowskian or Euclidean one according
to the choices of an ordinary or phantom model for the scalar
field. Here, after a brief review of the issue of the Noether
symmetry in a dynamical system we have applied the formalism into
a FRW cosmological model with a scalar field as its source. The
phase space was then constructed by taking the scale factor $a(t)$
and scalar field $\phi(t)$ as the independent dynamical variables.
The Lagrangian of the model in the configuration space spanned by
$\left\{a, \phi\right\}$ is so constructed such that its variation
with respect to these dynamical variables yields the appropriate
field equations. As is mentioned above, we introduced a new set of
variables $(x,y)$ in terms of which the minisuperspace takes the
form of a Minkowskian space, in the case of a ordinary scalar
field, and a Euclidean space in the case where the scalar field is
of phantom type. Also, the dynamics of the $(x,y)$ system is
described by a Lagrangian which is a diagonal quadratic form with
constant coefficients. The existence of Noether symmetry implies
that the Lie derivative of this Lagrangian with respect to the
infinitesimal generator of the desired symmetry vanishes. By
applying this condition to the Lagrangian of the model, we
obtained the explicit form of the corresponding potential
function. In the both cases of ordinary or phantom scalar field we
found three distinct solutions for the potential function each of
which have their own classical and quantum dynamics. In more
details , in the case where the scalar is an usual one the three
forms of Noether symmetric potentials are found as
$V(y/x)=y^2/(y^2-x^2)$, $V(y/x)=x^2/(x^2-y^2)$ and
$V=\mbox{cons.}=1$. In the first two cases one of the variables
has a cyclic dynamics while the other behaves linearly as time
progresses. Also, for a constant potential function, we saw that
the system of the two variables $(x,y)$ is a two dimensional
oscillator-ghost-oscillator system in which both of the variables
oscillate with the same frequency. The classical trajectories of
these models are also obtained and it is seen that they all
represent a bounded motion in the configuration space. We have
shown that all of these classical solutions have a singularity of
the big-bang or big-crunch type and so to pass this issue we have
dealt with the quantization of the model. As for the quantum
version of these models, we obtained exact solutions of the WDW
equation. The wave function of the corresponding universe consists
of some branches where each may be interpreted as part of the
classical trajectory. We saw that since the peaks of the wave
function follow the classical trajectory, there seems to be good
correlations between the corresponding classical and quantum
cosmology. The same study for a phantom scalar field yields the
potentials $V(y/x)=y^2/(x^2+y^2)$, $V(y/x)=x^2/(x^2+y^2)$ and
$V=1$. For these potentials, we solved the classical equations of
motion exactly and showed that at least one of the variables
behaves hyperbolically. This causes the classical trajectory in
$x-y$ plane to be an unbounded path in which the moving particle
(universe) tends to regions where the scale factor is large. The
resulting quantum cosmology and the corresponding WDW equation in
the phantom framework were also studied and analytical expressions
for the wave functions of the universe were presented with good
correlations with the classical trajectories.

Finally, we would like to emphasize that for an ordinary scalar
field the quantum effects dominate in the region of the classical
big-bang singularity, i.e., at the small values of scale factor.
At the big-bang the quantum solutions bounce from a contraction
phase to an expansion era. On the other hand for a phantom field
the quantum effects dominate in the region of the classical large
scale factor so that in this region quantum solutions fall from an
expansion phase to a contraction era.

\end{document}